\pdfoutput=1
\RequirePackage{amsmath}
\documentclass{svproc}
\usepackage{cleveref}
\usepackage{tabularx}
\usepackage{subcaption}
\usepackage{xspace}

\usepackage{cite}

\usepackage{breqn}

\newcommand{\etal}{\emph{et~al.}\xspace}
\newcommand{\highschool}{\textbf{HighSchool}\xspace}
\newcommand{\usflights}{{\bf US-Flights}}
\usepackage{tikz}
\usepackage{pgfplots}
\usepackage{amsmath}
\usetikzlibrary{decorations.pathmorphing, positioning}
\definecolor{echoreg}{HTML}{2cb1e1}
\definecolor{echodrk}{HTML}{0099cc}
\tikzstyle{mybox} = [text=black, very thick,
    rectangle, rounded corners, inner sep=10pt, inner ysep=20pt]
\tikzstyle{fancytitle} =[text=black]

\tikzstyle{every node}=[font=\scriptsize]

\tikzset{RootStyle/.style = {
						shape          = circle,
			      draw           = red!50!black!50,
			      thick,
			      top color      = white,
			      bottom color   = red!50!black!20,
			      text           = black,
			      inner sep      = .2pt,
			      outer sep      = 0pt,
			      minimum size   = 2.5 mm}
						}

\tikzset{VertexStyle/.style = {
						shape          = circle,
			      draw           = black!50,
			      top color      = white,
			      bottom color   = black!20,
			      text           = black,
			      inner sep      = .2pt,
			      outer sep      = 0pt,
			      minimum size   = 8 mm}
						}

\tikzset{EdgeStyle/.style   = {thick,-}}

\tikzset{LabelStyle/.style =   {
				  text           = black,
				  inner sep      = .2pt,
				  outer sep      = 1pt,
				  font           =\scriptsize,
				  minimum size   = 2.15 mm}
					}

\pgfplotsset{compat=1.14}

\usepackage{url}

\begin{document}
\mainmatter              
%
\title{Introducing multilayer stream graphs \\and layer centralities\vspace{-.2cm}}
%
\titlerunning{Introducing multilayer stream graphs and layer centralities}
%
\author{P. Parmentier$^{1 2}$, T. Viard$^1$, B. Renoust$^{3 4}$, J-F. Baffier$^{1 5}$}
\authorrunning{Pimprenelle Parmentier \etal} 
%
\tocauthor{P. Parmentier$^{1,2}$, T. Viard$^1$, B. Renoust$^{3,4}$, J-F. Baffier$^{1,5}$}
\institute{$^1$RIKEN AIP, $^2$\'Ecole Polytechnique, 
%
%
$^3$Institute for Datability Science, Osaka University, 
$^4$JFLI, CNRS UMI3527,
$^5$Japan Society for the Promotion of Sciences\vspace{-.3cm}
}

\maketitle              

\begin{abstract}
Graphs are commonly used in mathematics to represent some relationships between items. However, as simple objects, they sometimes fail to capture all relevant aspects of real-world data. To address this problem, we generalize them and model interactions over time with multilayer structure. We build and test several \emph{centralities} to assess the importance of layers of such structures. In order to showcase the relevance of this new model with centralities, we give examples on two large-scale datasets of interactions, involving individuals and flights, and show that we are able to explain subtle behaviour patterns in both cases.

\keywords{\vspace{-.2cm}multilayer graph, stream graph, centrality, density\vspace{-.1cm}}
\end{abstract}

\section{Introduction}
\label{sec:introduction}
\vspace{-.2cm}

Graphs have been widely used since the first definition of the K\"{o}nigsberg Bridges by Euler~\cite{euler1741solutio}. Although their formalization and drawing came later~\cite{hopkins2004truth}, graphs have been constantly challenged and their formalism extended in many ways, with, among the most common, orientation, labels, and weights for nodes and links~\cite{biggs1986graph}, to represents the connections of things in their entirety: friendships, railroads, communications, \emph{etc.}

Recently, new formalisms have emerged to encompass the more complex patterns that arise in real-world data. In particular, the multilayer networks~\cite{kivela2014multilayer} capture multiple families of relationships and entities together. This is useful, for example, to inspect homophily within groups of documents~\cite{renoust2014entanglement}. However, multilayer networks show some limitations in fully capturing interactions that exist over time~\cite{renoust2017multiplex}, beyond the dynamic of a graph as a whole. To cope with many individual time-dependant interactions, stream graphs~\cite{latapy2018stream} offer a comprehensive formalism to deal with real-world sequences of interactions over time.

In this paper, we are interested in joining both formalisms by proposing the \emph{multilayer stream graph}. After briefly reviewing the state of the art, we give its definition in \Cref{sec:formalism}, we explore the notion of centrality in \Cref{sec:centralities}, while applying this model on two datasets in \Cref{sec:usecases} before concluding. The key contributions of this paper are (i) the introduction of multilayer stream graphs and (ii) the demonstration of its relevance for the detection of central layers. 


\vspace{-.2cm}
\section{Related Work}
\label{sec:relatedwork}
\vspace{-.2cm}
We now introduce the concept of multilayer stream graph. To illustrate our definitions, we will use a toy example of a population of monkeys $F_1, F_2, M_1, M_2$ (two females and two males) interacting together.

\subsection{From graphs to multilayer and  stream graphs}

A simple graph is a tuple $G=(V,E)$ composed by a set of nodes $V$ and a set of edges $E\subseteq V \otimes V$, where each edge is an unordered pair of two distinct nodes.
The {\em degree} of a node $v\in V$ $d(v)$ is the number of edges in which $v$ appears: $d(v)=|\{(u,w) \in E \subseteq V \otimes V| u=v \cup w=v\}|$.
The {\em density} is the probability, given two nodes, that they are connected: $\delta(G)= \frac{2|E|}{|V|(|V|-1)}$.
In \Cref{fig:monkeys}a, the nodes are monkeys: $V=\{F_1, F_2, M_1, M_2\}$ and there is a link in $E$ between two animals if they have been in contact, hence forming $G=(V,E)$. $M_1$ is the monkey with the highest degree. In this example, the density $\delta(G)=\frac{2}{3}$.

\subsubsection{Multilayer graphs}
\begin{figure}[t!]
    \centering
    \begin{subfigure}[b]{.20\textwidth}
			\centering
            \includegraphics[width=\textwidth]{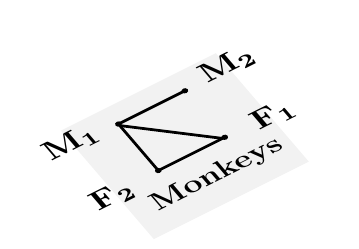}
            \vspace{0.35cm}
			\caption{Simple graph}
			\label{subfig:monkeysimple}
	\end{subfigure}
        \begin{subfigure}[b]{.30\textwidth}
			\centering
            \includegraphics[width=\textwidth]{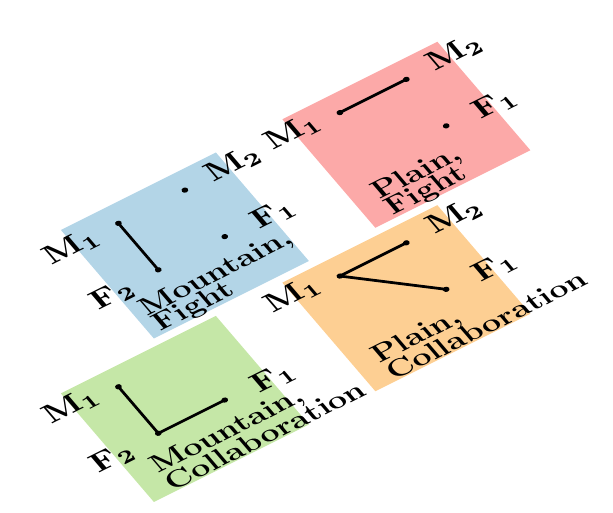}
			\caption{Multilayer graph}
			\label{subfig:monkeyplex}
    \end{subfigure}
		\hfill
    \begin{subfigure}[b]{0.47\textwidth}
			\centering
        \includegraphics[width=\textwidth]{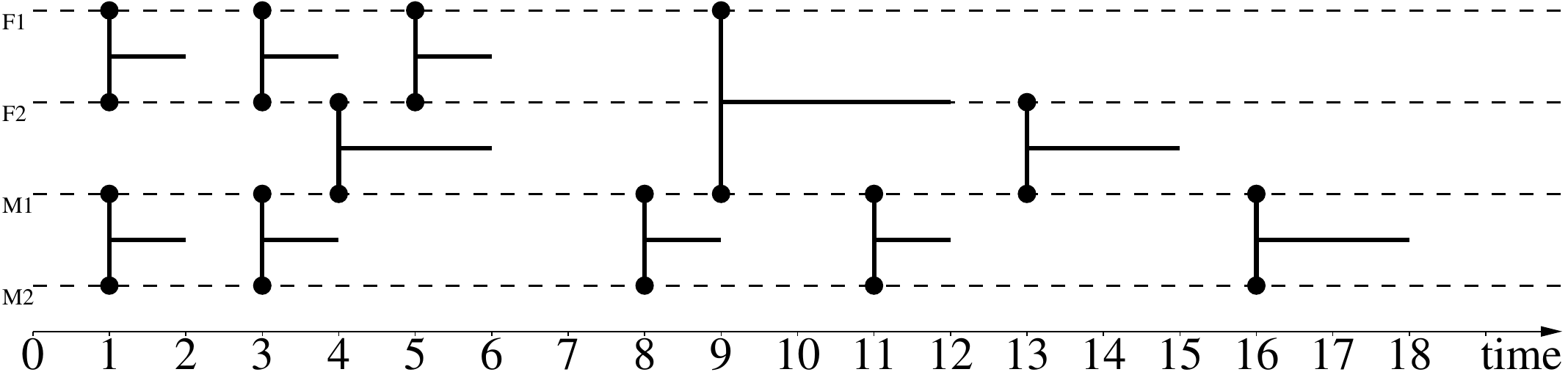}
    \vspace{0.35cm}
    \caption{Stream graph}
			\label{subfig:monkeystream}
    \end{subfigure}
    \caption{\footnotesize Example of the interactions of a population of monkeys. (a) The simple social network of monkeys. (b) The multilayer graph between monkeys across places and relationships (collaborating and fighting in the mountain or in the plain). (c) The stream graph describing sequence of interactions between monkeys. We may notice the frequent interactions between monkeys.}
    \label{fig:monkeys}
\end{figure}

Consider now 
that the monkeys from \Cref{fig:monkeys}a to interact in different places, such as the \emph{mountain} or the \emph{plain}, 
either through \emph{collaboration} or \emph{fight}.

A {\em multilayer graph}~\cite{kivela2014multilayer,de2013mathematical} is a set $M=(V_M,E_M,V,{\cal L})$, where ${\cal L}$ is the {\em structure}, a finite set of $d$ different sets, named {\em aspects} such that ${\cal L} = \lbrace L_1, \dots, L_d\rbrace$. Each aspect $L_i$ contains elements $l_i^1,\dots l_i^{n_i}$ which are named {\em elementary layers}. A {\em layer} $\alpha$ is then a combination of elementary layers from each aspect: $\alpha \in L=L_1\times \dots \times L_k$. $V$ is the set of nodes and each of them can be present on any of the different layers. A node on a layer is called a {\em node-layer}. The set of nodes-layers is $V_M \subseteq V \times L$. 
Each node-layer can be linked to another with undirected edges, \emph{i.e.}\ $E_M\subseteq V_M \otimes V_M$. 
The {\em degree of a node-layer} is the number of links in which the node-layer appears. The {\em degree of a node} is the number of links in which the node appears. The {\em density} can also be defined for each layer, hence generalized over the multilayer network: $\delta(M)= \frac{2|E_M|}{|V_M|(|V_M|-1)}$. 

\Cref{subfig:monkeyplex} represents our monkeys example in a multilayer context.
The structure ${\cal L}$ is made of two aspects: the place and the type of interaction. Consider the layer \emph{(mountain,collaboration)} in which a link between $M_1$ and $F_2$ means that the two animals collaborated in the mountain. 
The layer \emph{(mountain,collaboration)} and \emph{(plain,collaboration)} show higher density than the other layers. A possible interpretation is that this group of monkeys interactions are more collaboration based. Also, $F_2$,\emph{(mountain,collaboration)} and $M_1$,\emph{(plain,collaboration)} are the two key individuals with the highest collaboration degree.


\subsubsection{Stream graphs}

However, this does not take into account the temporal nature of interactions, {\em i.e.} when and for how long they occur. For example, fights or collaborations could be short or long depending on the circumstances, and occur frequently or not.
This information is crucial for finer grained understanding.

A {\em stream graph}~\cite{latapy2018stream} is a tuple $S=(T,W,V,E)$ where
$T$ is the time interval of study, $V$ is the set of nodes.
The stream graph model does not require a discrete definition of time. A stream graph considers a set of \emph{time instants} in a continuous manner.

The {\em time-nodes} set $W \subseteq T \times V$ describes the existence of nodes depending on time: $(t,v) \in W$ means that the node $v$ appears at time instant $t$. The set $E \subseteq T \times V \times V$ contain all the links and their time instants of existence. Given nodes $u$ and $v$, we call $T_u = \lbrace t, (t,u) \in E\rbrace$ the \emph{set of time instants} at which $u$ appears, and $T_{uv} = \lbrace t, (t,uv)\in E\rbrace$ the set of time instants at which the link $(u,v)$ appears.


Several notions have been designed in~\cite{latapy2018stream} that extend the model of classical graphs. 
For example, links may either last a certain amount of time, or  be instantaneous (leading to a density equal to zero). 

The {\em number of links} of a set of links is formally define as the duration of the links divided by the length of $T$. The {\em degree} of one node is the {\em number of links} of the set of links attached to the node.
The {\em density} $\delta(S)$ is the probability, given two nodes and a time instant, that a link exists between the two nodes:
\[ \delta(S) = \frac{\sum\limits_{e \in E} |T_e|}{\sum\limits_{u,v \in V\otimes V} |T_u \cap T_v|}\]

Figure~\ref{subfig:monkeystream} 
provides a new distinctive look at the interactions happening between the individuals over 10 units of time. 
$M_1$ is the node with the highest degree of value $0.55$. We can also observe that $F_1$ and $F_2$ met twice shortly. $F_1$ and $M_1$ are the first to meet. The density of the stream graph is $\delta(S) = \frac{16}{20 \cdot 4 \cdot 3}\approx 0.06$.

\subsection{Temporality, multiple layers, and centrality}

Now that we have introduced both the multilayer graph and stream graph models, let us briefly review prior works relevant to both temporal and multilayer approaches.

We may first be interested by the dynamics in multilayer networks~\cite{bassett2013robust, kivela2014multilayer, gallotti2015multilayer, de2016physics, amelio2017evolutionary, pilosof2017multilayer, renoust2017multiplex, mourchid2018multilayer}. In temporal multilayer networks, one goal is often to identify community structures~\cite{bassett2013robust, amelio2017evolutionary, mourchid2018multilayer}, which is not the task we are focusing on this paper. 
However, to model such networks, Kivel\"{a} \etal~\cite{kivela2014multilayer} suggest that temporality is only an \emph{aspect} of the multilayer networks that could decompose the multilayer network like any other aspect. The same approach is taken by Pilosof \etal~\cite{pilosof2017multilayer}. Nonetheless, unlike any aspect, time is submitted to order. Moreover, these analyses only consider juxtaposed time-frames as separated networks. Considering the changes of topologies in an overall graph is however relevant for the study of spreading processes~\cite{gallotti2015multilayer, de2016physics}, which demonstrated the dependency on spreading from layers coupling. These works also point out how the evolution of centrality is key to studying the network~\cite{taylor2017eigenvector}, and isolating nodes of interest, such as in citation networks~\cite{renoust2017multiplex}.

As of heterogeneity in stream graphs, we may first bring forward the $\Delta$-analysis~\cite{latapy2018stream}, which provides a way of studying interactions at multiple time scales; it may be regarded as a specific case of multilayer.
Some approaches are more hybrid between the two. Vaiana \etal~\cite{vaiana2018multilayer} are the first to bring a hybrid model between dynamic multilayer and temporal stream of links to capture different functional networks in the brain, but they mostly use the multilayer dynamic approach and introduced a unifying definition as a future work, which is in line with our contribution.

Both dynamic multilayer graph models or stream graphs would work with temporal data, however each imposes a specific point of view on the data.
Each time frame in a dynamic multilayer network imposes to choose a time granularity, the choosing of which is not trivial. In addition, it also implies some distortions: either parts of links duration would be excluded (outside of the time frame) or a link would be considered as present over the whole frame considered. Multilayer stream-graphs, our proposed model, are completely agnostic to these issues, since they take each link in its own duration. The focus of this model is not on a whole graph interacting, but closer to the data, on a series of interaction events, no matter the nodes they attach. The structure of the resulting graph is only a consequence of these interactions.

\section{Multilayer stream graph}
\label{sec:formalism}

Let us now introduce our object. A {\em multilayer stream graph} $S_M$ is a tuple $(T, V, {\cal L}, L_M, V_M, W_M, E_M)$, with $T$ a time interval, $V$ a set of nodes, and ${\cal L} = \{{\cal L}_i\}_{i=1}^d$ a set of {\em aspects}.
For a given $i\leq d$, $\epsilon \in {\cal L}_i$ is called an {\em elementary layer}; finally, we call a {\em layer} an element of $L={\cal L}_1 \times {\cal L}_2\times \dots \times {\cal L}_d$.
We denote by $V_M \subseteq V \times L$ the set of node layers, and by $W_M\subseteq T\times V \times {\cal L}$ the set of time-nodes-layers. In other words, $(u, \alpha)\in V_M$ means that node $u$ is present in layer $\alpha$, and $(t,u,\alpha)\in W_M$ means that node $u$ is present at time $t$ in layer $\alpha$. 
$L_M \subseteq I\times L$ is the set of time-layers, where $I$ is the set of intervals included in $T$. $(t,\alpha)$ in $L_M$ means that layer $\alpha$ exists at time $t$.
Finally, we denote by $E_M \subseteq T\times V_M \otimes V_M$ the set of interactions. In other words, $(t, (u,\alpha), (v,\alpha') )\in E_M$ means that node $u$ in layer $\alpha$ and node $v$ in layer $\alpha'$ interacted at time $t$.

We illustrate this object in \Cref{subfig:monkeyplex}. Notice that $(t,((u,\alpha),(v,\beta))) \in E_M $ implies that $(t,u\alpha) \in W_M$ and $(t,v,\beta) \in W_M$. Similarly, $(t,u,\alpha)\in W_M$ implies that $(t,\alpha) \in L_M$. In other words, a link between two nodes-layers can only exist if the two nodes-layers exist at this time, and nodes-layers can only be present when layers exist.


    \begin{figure}[t!]
    \centering
    	\begin{subfigure}[b]{0.35\textwidth}
            \centering
        	\includegraphics[width=\textwidth]{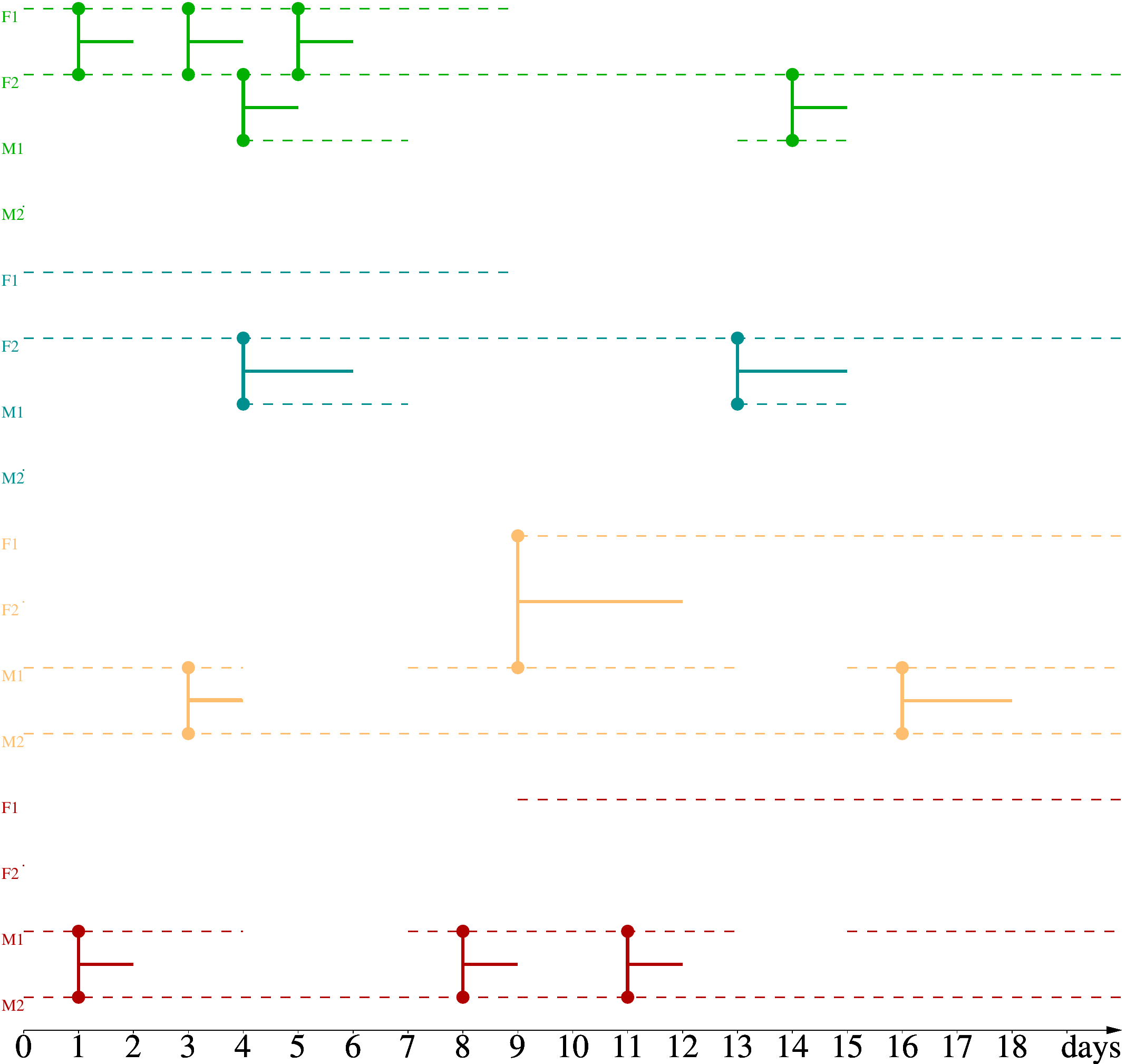}
        	\caption{Monkeys example}
    	\label{fig:monkeymls}
    	\end{subfigure}
    	\begin{subfigure}[b]{0.60\textwidth}
            \centering
        	\includegraphics[width=\textwidth]{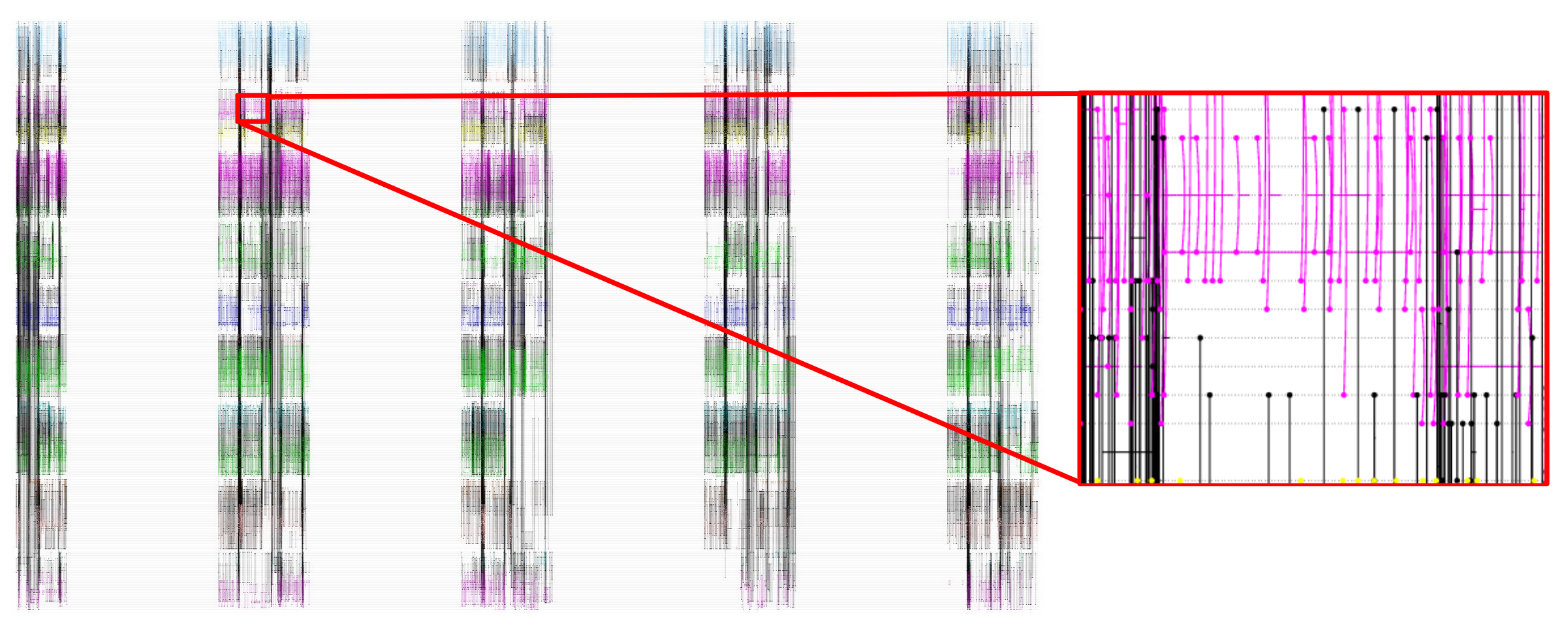}
        	\vspace{0.35cm}
        	\caption{Real example: \highschool dataset}
            \label{fig:visuLycee}
        \end{subfigure}
    \caption{(a) 
        	This example follows the colors of \Cref{subfig:monkeyplex}: in green, {\em (mountain,collaboration)}; {\em (Fight,mountain)}, in blue; {\em (collaboration, plain)}, in yellow; and t{\em (fight,plain)}, in red. This example captures 13 interactions of the 4 monkeys in a time frame $[0,20]$. 
        	(b) The visualization of the much larger multilayer stream graph for dataset \highschool composed of $36 732$ links involving $329$ students over the course of $5$ days.}
        	    	
    \end{figure}

In the rest of this paper, for the sake of defining the most elementary object possible, we consider multilayer stream graphs to be undirected, unweighted and unlabeled.
We show that even this elementary model is relevant for real-world data analysis.
We discuss some possible extensions in \Cref{sec:conclusion}.

\bigskip
Let us now start by defining various extractions and projections of multilayer stream graphs.


The {\em induced multilayer graph} by the set $\tau \subseteq T$ is a multilayer graph $M_I(S_M)=(V_{M,I}, E_{M,I}, V,L)$ which gathers all the layers, nodes-layers and links existing over time $\tau$. In this graph, $V_{M,I}  = \{ v \in V | \exists t \in \tau, (t,v) \in V_M\}$ and $E_{M,I}  = \{((v,\alpha),(w,\beta)) \in (V\times L)\otimes (V\times L) | \exists t \in \tau , (t,v,\alpha,w,\beta) \in E_M \}$.
In other words, it is the graph where nodes are elements of $V$ and one puts a link between two nodes if and only if they have interacted over a duration $\tau$.

Notice that when $\tau={t}$, this induced graph is called the {\em multilayer graph at time} $t$.
%
\Cref{subfig:monkeyplex} shows the multilayer graph induced by $T$ for the multilayer stream graph in \Cref{fig:monkeymls}.



Interlayer links are often used to model transit in a multilayer, such as the underground path to change between two lines of a subway station~\cite{kivela2014multilayer}. 
Given a pair of layers $\alpha, \beta \in L_1\times \dots\times L_d$, the {\em interlayer stream graph} is the bipartite stream graph $S^{(\alpha,\beta)} = (T^{\alpha,\beta}, V^{\alpha,\beta},W^{\alpha,\beta},E^{\alpha,\beta})$, with $T^{\alpha,\beta}=\{t: \exists (t,u,\alpha)\in W_M, \exists (t,u,\beta)\in W_M, u \in V\}$ the interval of time in which $\alpha$ and $\beta$ appear simultaneously. $V^{\alpha,\beta} = (V\times \{\alpha, \beta\})\cap V_M$ is the set of node-layers of the multilayer stream graph restricted to $\alpha$ and $\beta$. $W^{\alpha,\beta} = (T^{\alpha, \beta}\times V^{\alpha,\beta})\cap W_M$ describes all their intervals of existence. Finally, $E^{\alpha,\beta} =
\{(t, (u, \alpha)(v, \beta))\in E_M : t\in T^{\alpha, \beta}, u,v\in V\}$ is the set of interactions between layers $\alpha$ and $\beta$. 



For a given layer $\alpha$, we define the {\em intralayer stream graph} simply as $S(\alpha,\alpha)$, {\em i.e.} the interlayer stream graph between layer $\alpha$ and $\alpha$.
We denote it by $S^{\alpha} = (T^\alpha, V^\alpha, W^\alpha, E^\alpha)$. 
For example, \Cref{fig:monkeymls} captures only intralayer interactions.

    



The {\em aggregated stream graph}  $S_A(S_M)=(T,V,W_A,E_A)$ is the stream graph where all layer information has been removed. As such, it has the same interval of study $T$ as $S_M$. Its nodes are the same as in $S_M$ (the set $V$). Their times of existence are the union of their times of existence on the different layers: $T_u = \bigcup_{\alpha \in L} T_{u,\alpha}$ and $W_A=\bigcup_{u\in V} T_u\times\{u\}$. An edge exists between two nodes of $S_A(S_M)$ if it exists a the same time between two correspondent node-layers of $S_M$, {\em i.e.}  $E_A = \{(t,u,v)| \exists (\alpha,\beta) \in L^2, (t,(u,\alpha),(v,\beta)) \in E_M \}$. For example, \Cref{subfig:monkeystream} is the aggregated stream graph obtained by superimposing all layers in \Cref{fig:monkeymls}.


The {\em degree} of a node in a multilayer stream graph is the number of links (as defined just before) in which the node appears, {\em i.e.} $d(u) = |\{(t, (u,\alpha)(v, \beta))\in E_M : t\in T, v\in V, \alpha, \beta \in {\cal L}\}|$. Similarly, the degree of a node-layer $(u,\alpha)$ is simply $d(u, \alpha) = |\{(t, (u,\alpha)(v, \beta))\in E_M : t\in T, (v, \beta) \in L\}|$.

In \Cref{fig:monkeymls}, we can notice that females interact much more in the mountain than in the plain,and the contrary for the males. We can spot that the longest interaction, between $F_1$ and $M_1$, takes place in the plain and lasts 3 days.
The node with the highest degree is $M_1$ ($d=7$) and the node-layer with the highest degree is ($M_1,$\emph{(plain,collaboration)}).


The {\em density} of a multilayer stream graph is the probability, when one takes a random time $t$ and two random node-layers $(u,\alpha)$ and $(v,\beta)$ that the link $(t,(u,\alpha),(v,\beta))$ is in $E_M$:
\(
	    \delta(S_M)
	    = \frac{\sum\limits_{(u,\alpha),(v,\beta) \in E_M}|T_{(u,\alpha)(v,\beta)}|}{\sum\limits_{(u, \alpha), (v, \beta)} |T_{(u,\alpha)} \cap T_{(v,\beta)}|}
\).

In \Cref{fig:monkeymls}, the density of the multilayer stream graph is: $\delta(S_M) = \frac{18}{104}\approx 0.17$. 
Notice that in comparison, the density of the aggregated stream graph is $\delta(S) = \frac{17}{20*6}\approx 0.14$, and the one of the aggregated graph is $\delta(G) = \frac{4}{6} = \frac{2}{3}$.

Moreover, this definition of density can be readily applied and adapted to specific cases.
For example, the interlayer density of interactions between two layers $\alpha$ and $\beta$ is nothing but $\delta(S_M(\alpha, \beta))$, the density of the interlayer stream graph. 
The denominator sum can also be modified to take into account specific aspects of the data, for example by summing on $(u, \alpha), (v, \alpha)$ if interlayer links are not allowed, among others.

\subsection{Centralities}
\label{sec:centralities}

One key application on real-world data is the analysis and detection of important nodes, {\em i.e.} that are {\em central}.
Many notions of centrality coexist for graphs~\cite{brin1998anatomy,brandes2001faster}, multilayer graphs~\cite{renoust2015detangler,ghalmane2019centrality} and stream graphs~\cite{latapy2018stream,costa2015timecentrality} alike.
As of today, no consensus emerges on a global centrality notion, as they all capture different notions of importance~\cite{gupta2016centrality}.

In this section, we develop upon the formalism introduced in \Cref{sec:formalism} and introduce two centrality definitions on multilayer stream graphs extending from entanglement~\cite{renoust2014entanglement,renoust2015detangler} and inspired by eigenvector centrality~\cite{gupta2016centrality}, taking into account the multifaceted nature of the object while remaining simply explainable and computationally efficient.

As a prerequisite, let us extend the definition of {\em paths} to multilayer streams graphs.
A path from $(t, u, \alpha)$ to $(t', v, \beta)$ is a sequence $(ti,(u_i,\alpha_i),(v_i,\beta_i))_{i=0}^{k}$ of elements of $E_M$ such that $(u_0, \alpha_0) = (u, \alpha)$, $(v_k, \alpha_k) = (v, \beta)$, $t_0 \geq t$, $t_k \leq t'$ and for all $i=0..k$,  $(u_{i+1},\alpha_{i+1}) = (v_i, \beta_i)$ and $t_{i+1} \geq t_i$.
A common variant, defined in~\cite{latapy2018stream}, are $\gamma$-paths, {\em i.e.} paths for which the condition 
$t_{i+1} \geq t_i$ becomes $t_{i+1} \geq t_i+\gamma$.
In other words, in $\gamma$-paths traversing an edge costs $\gamma$. This is especially useful for modelling transportation networks, as we will see in \Cref{sec:usecases}.


Let us now introduce the new notions of centrality, that we call {\em superimposed layer centrality} and {\em juxtaposed layer centrality}.
Both aim at giving an intuition of the importance of {\em layers} in the multiplex stream graph.

A group of layers is {\em superimposed} if each node can be present on each layer at any time (also referred as multiplex networks~\cite{kivela2014multilayer}).
In other words, saying that two layers are superimposed means that it is possible to have interlayer links between those layers.
This typically corresponds to layers describing diverse types of relationships across the same set of nodes. For instance, in \Cref{fig:monkeymls}, the layers \emph{(mountain,collaboration)} and \emph{(mountain,fight)} are superimposed.

Given a superimposed multilayer stream graph and a time $t\in T$, for all nodes $u\in V$ and all layers $\alpha_{i=0..k}$, let us compute $X_{\alpha_i}(t,u)$ as the probability that a random walker starting from node $u$ at time $t$ will cross a link involving layer $\alpha_i$.
One then obtains a $|V|\times k$ matrix corresponding to the relative importance of each layer $\alpha_i$ for each node $u$.

We define the {\em superimposed layer centrality} as the maximal eigenvalue of $\Sigma_X$, the matrix of covariances of all random walkers.
In $\Sigma_X$, each term of the matrix is computed as $X_{\alpha_i,\alpha_j} = E[ (X_{\alpha_i} - E[X_{\alpha_i}])(X_{\alpha_j} - E[X_{\alpha_j}]) ]$.
Intuitively, the eigenvalues of $\Sigma_X$ give a ranking of the layers by decreasing importance.
The maximal eigenvalue corresponds to the maximal variance for a linear combination of $X_{\alpha_i}$ corresponding to the eigenvector of $\Sigma_X$.








However, not all datasets contain meaningful superimposed layers. To this end, we define another notion of centrality, the {\em juxtaposed layer centrality}.

A group of layers is {\em juxtaposed} if each node can only be present in one layer. This is typically the case for non-superposed states: age, gender, class number, \textit{etc.} Notice however that the layer associated to each node can in principle change over time.
For example if one were to consider the aspect \emph{age=\{baby, child, adolescent, adult, elderly\}} in \Cref{fig:monkeymls},  the layers \emph{(child,mountain,collaboration)} and \emph{(adult,mountain,collaboration)} are juxtaposed.
In this case, studying the relations between layers is particularly relevant.

We then consider the interlayer density matrix $\Delta$, {\em i.e} for each pair $i,j$ of layers, one computes the interlayer density $\delta(S_M(\alpha_i, \alpha_j))$\footnote{With the case where $i=j$ simply returns the intralayer density $\delta(S_M(\alpha_i))$}.

From the multilayer stream graph displayed in \Cref{fig:monkeymls}, we can alternatively consider two layers of: {\em male} interactions, and {\em female} interactions.  The matrix of densities for these layers ({\em male} and {\em female}) is
$\left(\begin{smallmatrix}
1/5 & 1/7 \\
1/7 & 1/6
\end{smallmatrix}\right)$.
One can notice that in this (toy) example, interactions are stronger in the males than in the females, and that the female and the male interact less with their opposite gender.

The {\em juxtaposed layer centrality} then correspond for each layer to its entry in the eigenvector associated with the maximum eigenvalue of $\Delta$.
Notice that in both cases, the Perron-Frobenius theorem~\cite{perron1907theorie,frobenius1908matrizen} states that a irreducible non-negative matrix has a maximum positive eigenvalue with an eigenspace of range 1. In our case, we know that $\Delta$ is non-negative by definition of the densities and irreducible unless we can share the layers into different groups that do not interact together.

\section{Results}
\label{sec:usecases}

\begin{figure}[t]
    \begin{center}
    \begin{subfigure}[b]{0.32\textwidth}
			\centering
    \includegraphics[width=1\textwidth]{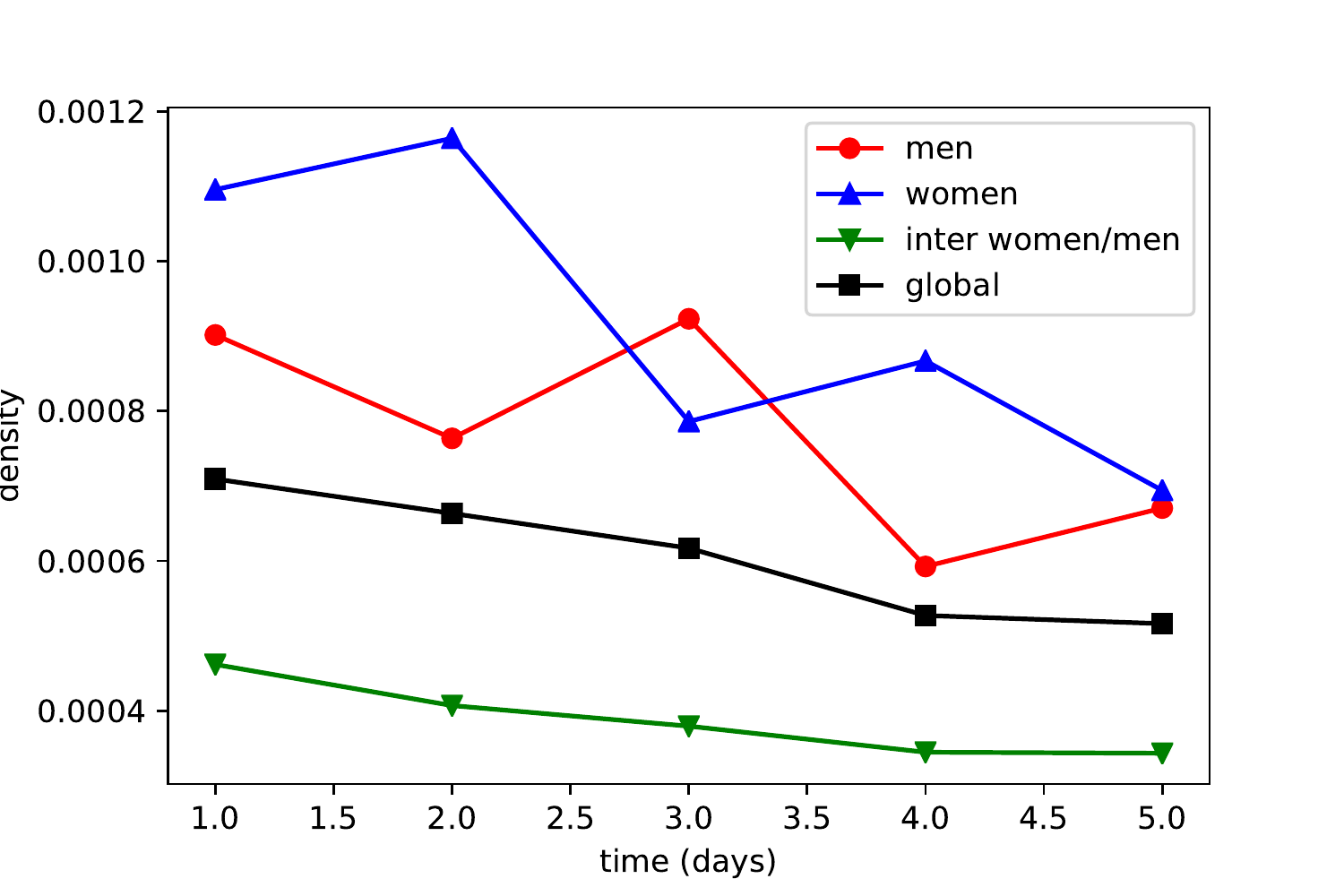}
    \caption{Density over time}
			\label{subfig:highschool-classes-densities-dyn}
    \end{subfigure}        \begin{subfigure}[b]{0.32\textwidth}
			\centering
	\includegraphics[width=1\textwidth]{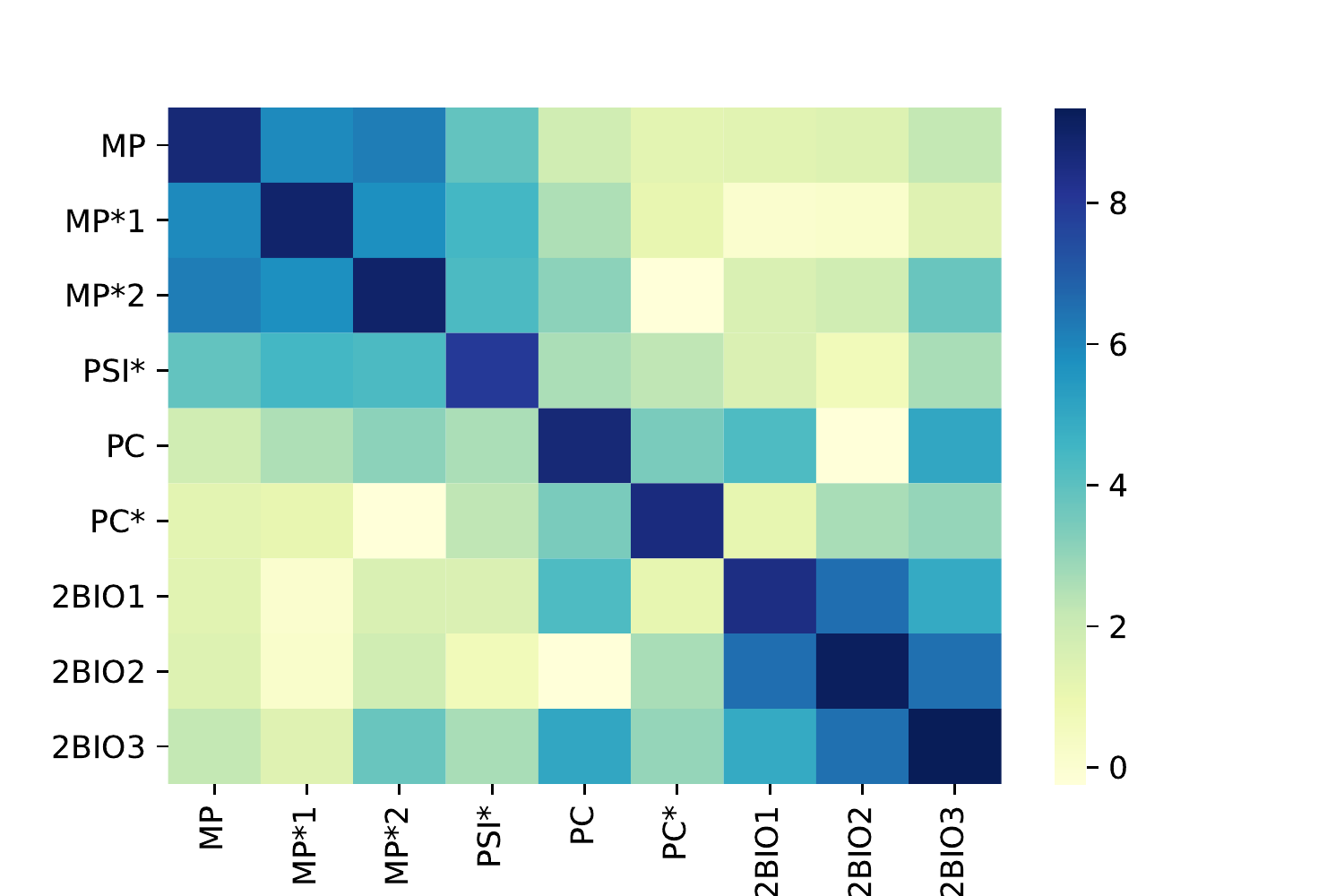}
    \caption{Densities}
			\label{subfig:highschool-classes-densities}
    \end{subfigure}    
    \begin{subfigure}[b]{0.32\textwidth}
			\centering
	\includegraphics[width=1\textwidth]{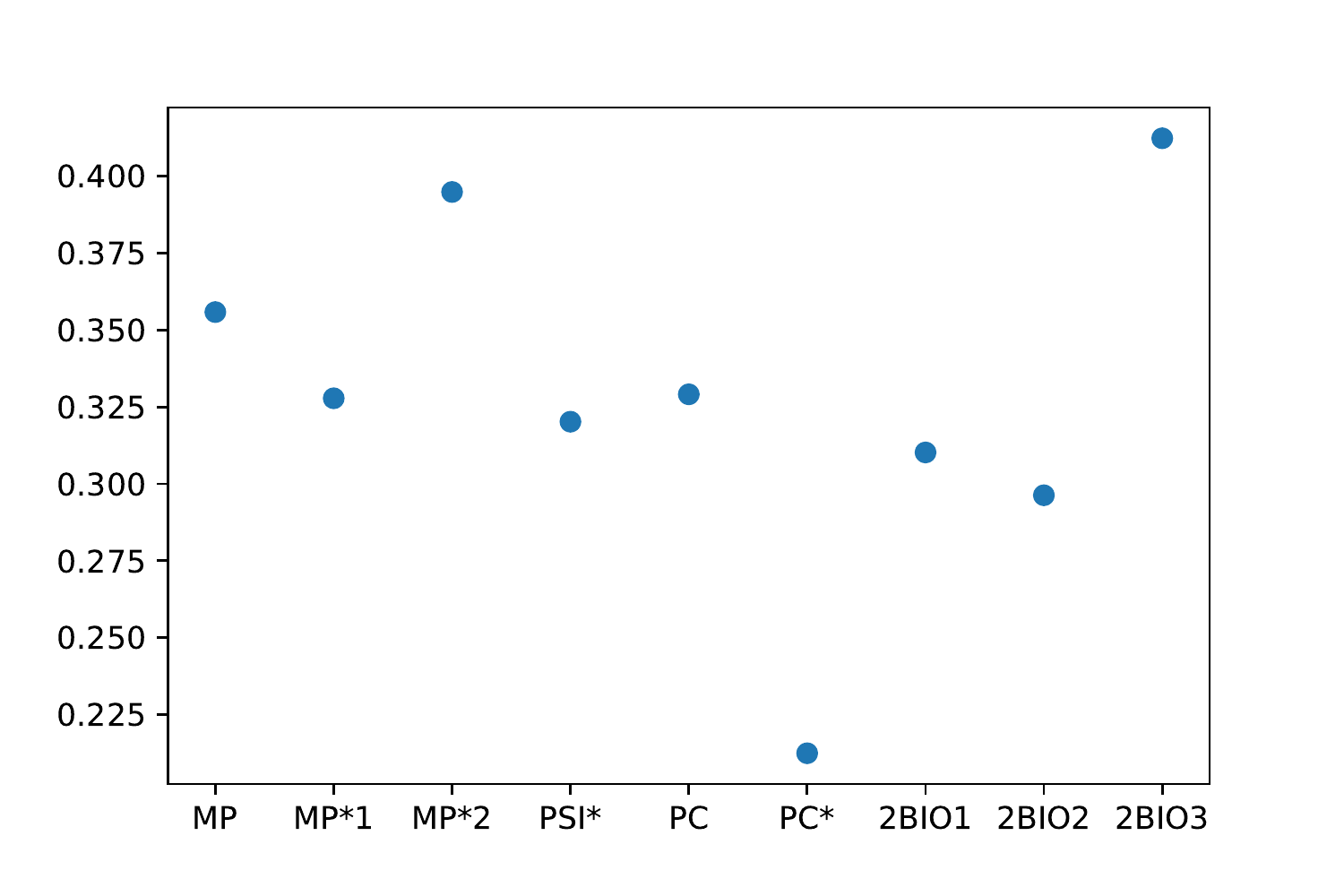}
    \caption{Centrality}
			\label{subfig:highschool-classes-centrality}
    \end{subfigure}    
    \caption{(a) Density in multilayer stream graph computed for each day, inside the layers of men, the layers of women, between the two groups of layers and inside the whole group. (b) The matrix of the log of densities between the classes. (c) Score of centrality for the different layers.}
    \label{fig:highschool-classes}
    \end{center}
\end{figure}

We now demonstrate multilayer stream graphs for the analysis of two real-world datasets.
All implementations are available to the public\footnote{\texttt{https://github.com/TiphaineV/multiplex-streams/src/visualisation}}.

\subsection{Data}
\label{subsec:data}

The first one records interactions among high-school students~\cite{mastrandrea2015contact}.
Each student is associated to a class, and interactions can be of three kinds: (i) face-to-face, (ii) self-declared friendship, and (iii) Facebook friendship.
Notice that only (i) is time-dependent, (ii) is directed, and (iii) is undirected.

This dataset comprises of $36,732$ links involving $329$ students over the course of $5$ days.
{\em Superimposed aspects} are the interaction type (face-to-face, friendship, Facebook friendship), whereas {\em juxtaposed aspects} are the gender of each student (female, male, or undefined) and the class.
The (relatively) small size of this dataset makes it visualizable, see \Cref{fig:visuLycee}. 
We will hereafter refer to this dataset as \highschool.

The second one documents all domestic flights in the United States since $1987$~\cite{BTS2019}.
Since the whole dataset is too large to be efficiently processed, we focus instead on a longitudinal study across the years, using all the flights in January 1988, 1995, 2010 and 2019.

Each of these datasets involves $346$ airports and contains roughly $500 000$ flights.
There are no \emph{juxtaposed aspects}, however the company operating the flight is a natural \emph{superimposed aspect}.
There is a maximum of $17$ companies.
Finally, while the stream graph is too large to be visualized, we display the induced graph over a map of the United States in \Cref{subfig:usflights-graph}.
We will hereafter refer to this dataset as \usflights.

\subsection{Experiments}

We now devise two experiments in order to shed light into some patterns present in the two datasets described in \Cref{subsec:data}.

The intrinsic structure of the two datasets allow us to demonstrate the relevance of both notions of centrality described in \Cref{sec:centralities}.
We show that our formalism is able to shed some light on patterns in both datasets, which in turn serves as a proof of concept for our work.


The \highschool\ dataset features gender information about the participants.
In this context, we investigate the interaction patterns between the participants, taking into account this information.

Let us consider the multilayer stream graph describing the interactions among students, the layers here being the gender ($\{M, F\}$\footnote{The dataset also contains a few \texttt{'U'}, for Undefined, that corresponds to interactions with teachers. We do not consider them in the rest of this work.} and the class label ($\{MP,$ $MP^*1, MP^*2, PSI^*, PC, PC^*, 2BIO1, 2BIO2, 2BIO3\}$), corresponding to usual French names for such schools.

Regarding gender, let us consider the multilayer stream graph induced by every $24$ hours.
We obtain $5$ such stream graphs, corresponding to the $5$ days of the dataset recording.
On each daily multilayer stream graph, we compute the inter- and intralayer densities, as well as the graph density ({\em i.e.} discarding all gender and temporal information). We show the result in \Cref{fig:highschool-classes}.
Several conclusions can be observed: first of all, the graph density (legend ``global'') does not adequately capture the subtleties in the data.
It averages the intra- and inter- densities, that are in reality following two different modes; in other words, individuals interact more with individuals of the same gender as them.

We also study the densities of interactions between the different classes.
\Cref{subfig:highschool-classes-densities-dyn} shows the density matrix for each class.
For readability, it displays the absolute value of the logarithm of the densities, as it makes blocks more apparent.
While as for gender, the intra-density is higher than the inter-density, we can discern larger blocks grouping layers together: $\{MP, MP^*1, MP^*2\}$ $\{2BIO1,2BIO2, 2BIO3\}$. These blocks correspond to specialty topics, as $MP$ corresponds to mathematics and physics, while $BIO$ corresponds to biology. This result is intuitive, and so this serves as an argument that our model captures the interaction subtleties in the data.


\Cref{fig:highschool-classes} shows the superimposed layer centrality values for each class, as defined in \Cref{sec:centralities}. 

$2BIO3$ and $MP^*2$ are the most central: we can see that they are the most central among two clusters of layers, regrouping the $MP$ classes and the $BIO$ classes. In terms of data analysis, this makes sense since these two groups correspond to common speciality subdivisions in the French system: favouring Biology (BIO) or Mathematics and Physics (MP).
Notice finally that the $PC$ (Physics and Chemistry) layers in this school are the least central; one can then assume that the students in these classes interact less in time with other classes, which comes as a surprise considering that one speciality, Physics, is shared with the $MP$ students.

\bigskip

Notice however that the \highschool\ dataset does not have juxtaposed layers that we can study.
In order to demonstrate the interest of the juxtaposed layer centrality, we focus instead on the \usflights\ dataset.

\begin{figure}[t]
\begin{subfigure}{0.165\textwidth}
	\includegraphics[width=\textwidth]{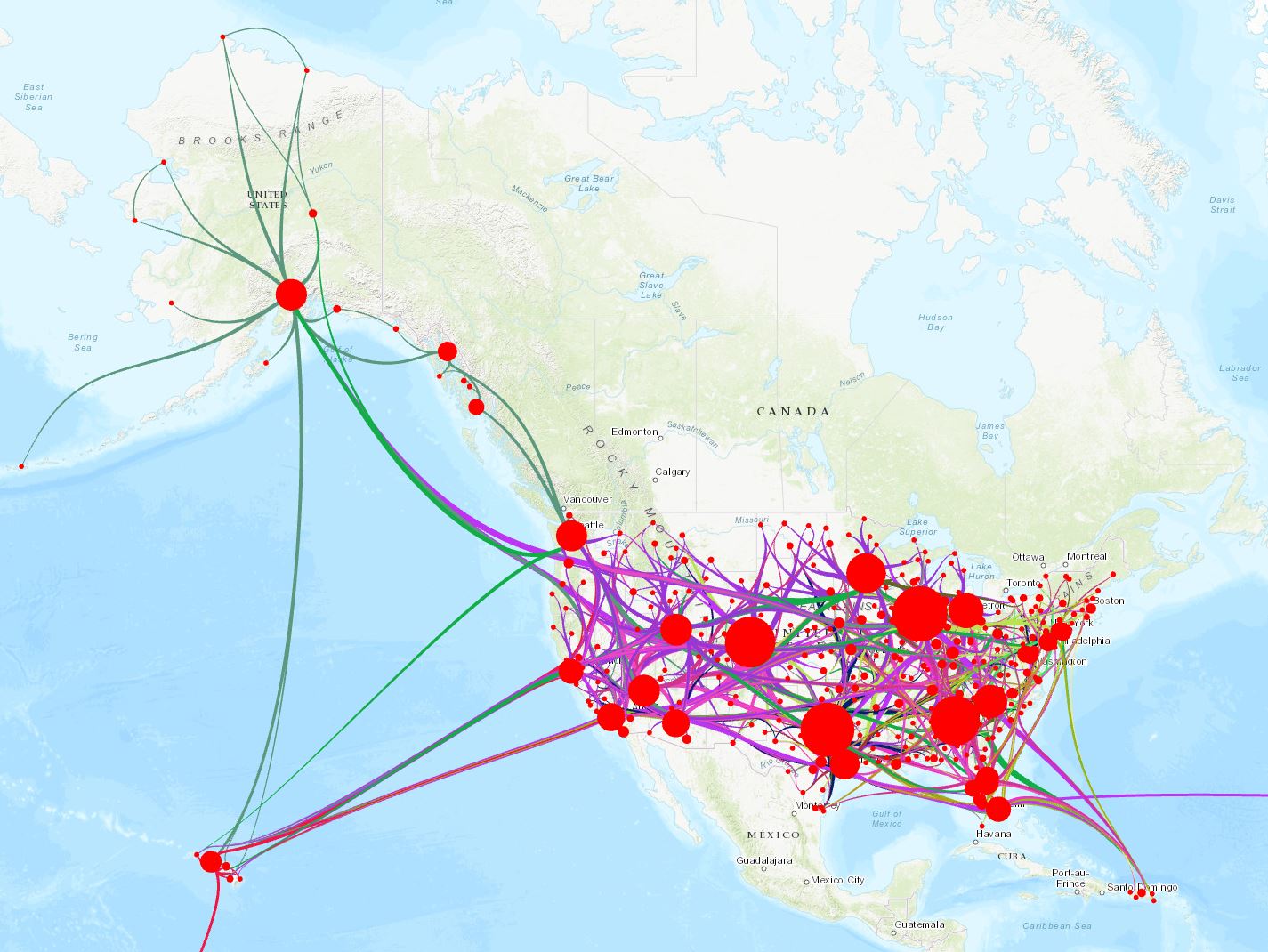}
	\caption{US-flights}
	\label{subfig:usflights-graph}
\end{subfigure}
\begin{subfigure}{0.19\textwidth}
	\includegraphics[width=\textwidth]{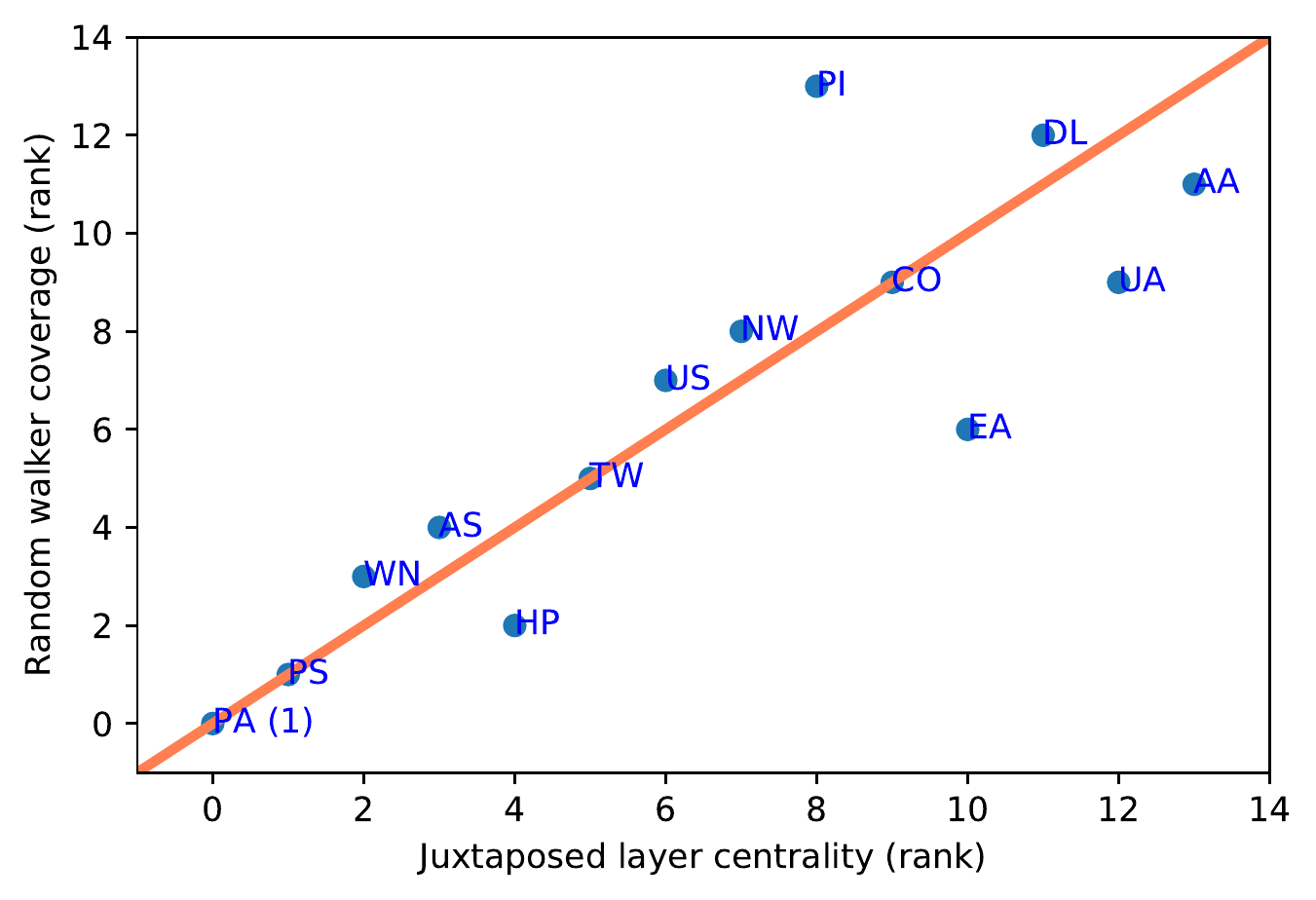}
	\caption{Jan. 1988}
	\label{subfig:rankplanes1988}
\end{subfigure}
\begin{subfigure}{0.19\textwidth}
	\includegraphics[width=\textwidth]{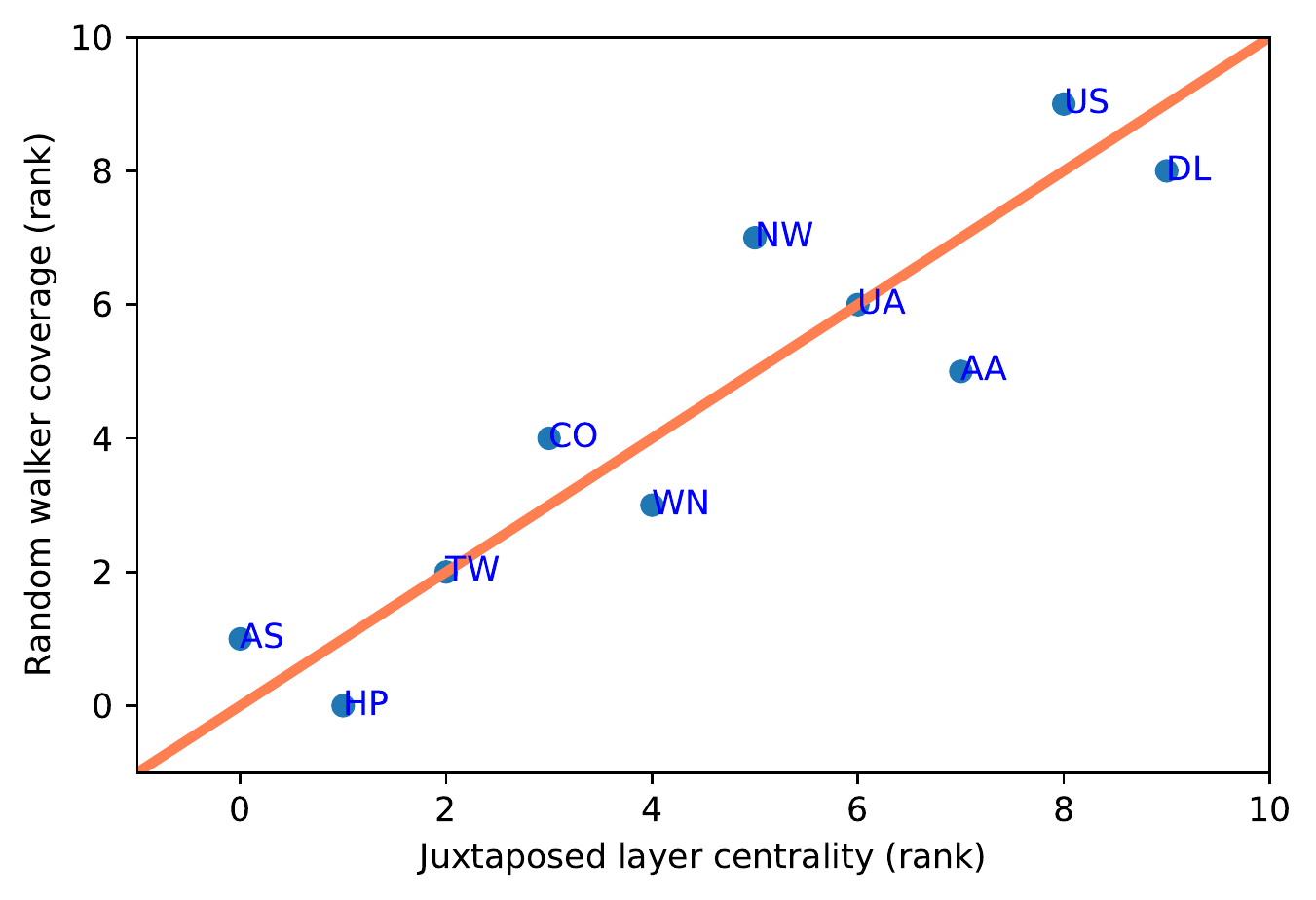}
	\caption{Jan. 1995}
	\label{subfig:rankplanes1995}
\end{subfigure}
\begin{subfigure}{0.19\textwidth}
	\includegraphics[width=\textwidth]{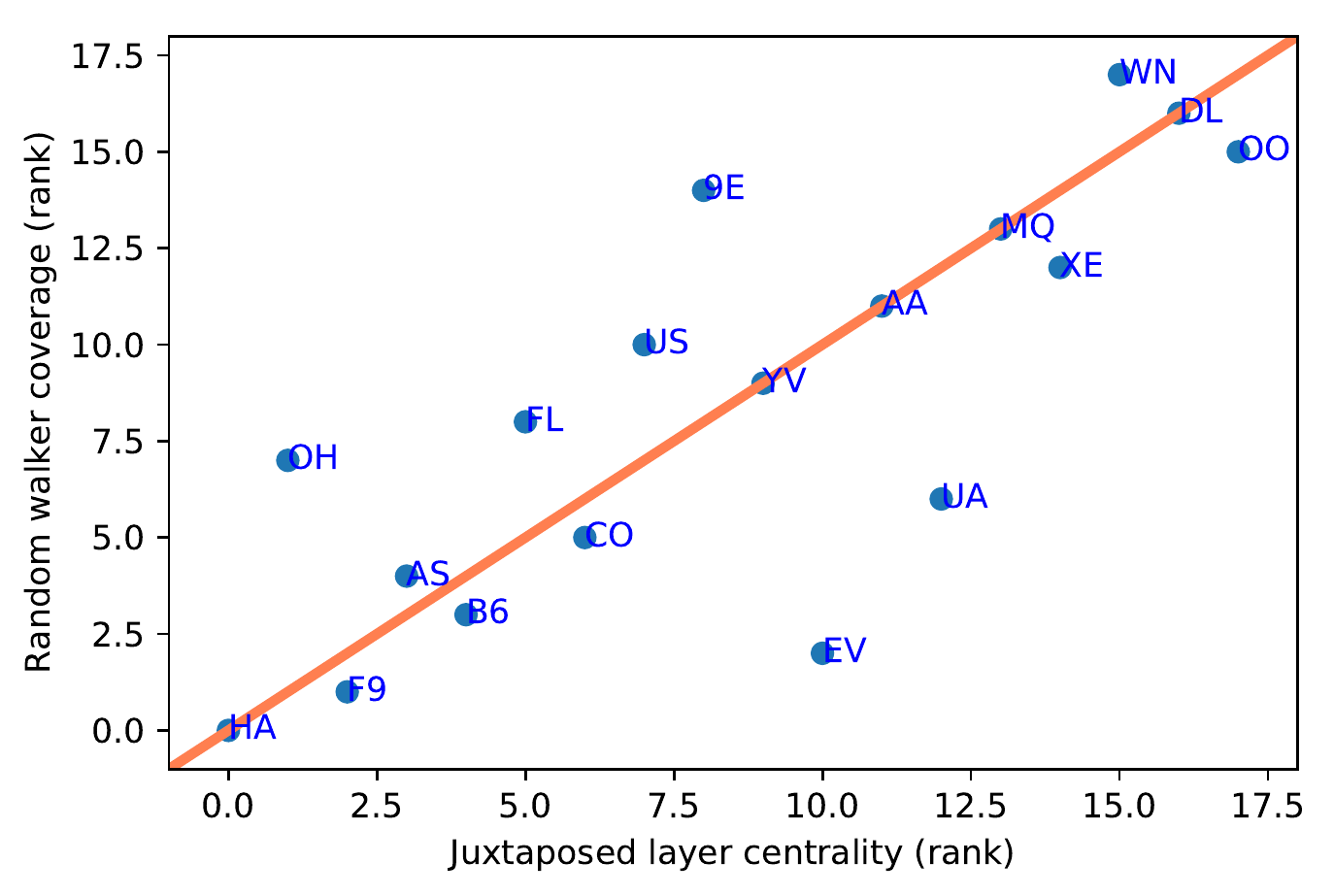}
	\caption{Jan. 2010}
	\label{subfig:rankplanes2010}
\end{subfigure}
\begin{subfigure}{0.19\textwidth}
	\includegraphics[width=\textwidth]{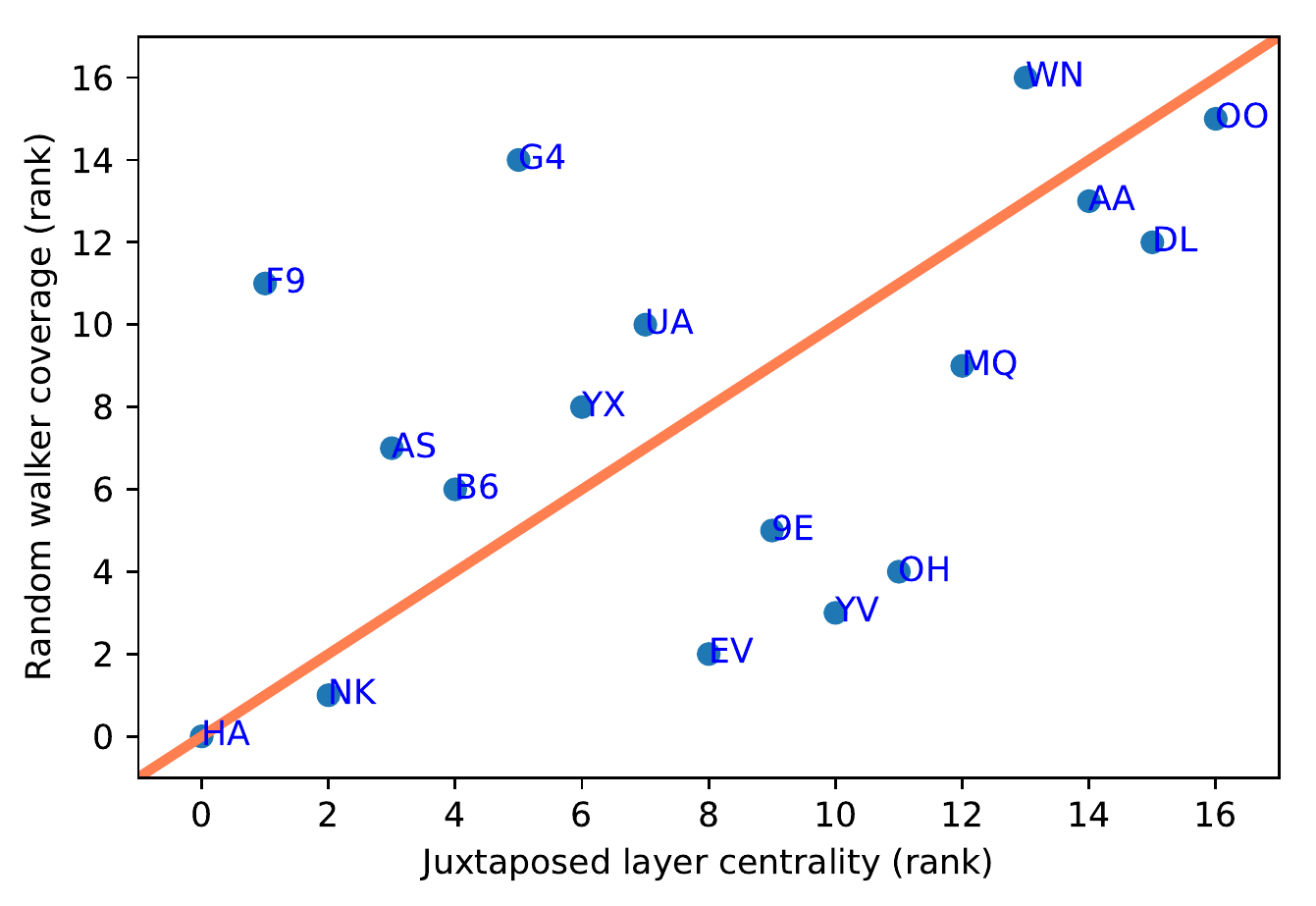}
	\caption{Jan. 2019}
	\label{subfig:rankplanes2019}
\end{subfigure}
\caption{(a) Induced multilayer graph of the US companies flights. The size of the nodes reflect of their PageRank. 
(b-e) Comparison between Rank by coverage and juxtaposed layer rank of companies for the $4$ subdatasets extracted from \usflights. In red, the function $y=x$, corresponding to a perfect correlation.}
\label{fig:rankplanes}
\end{figure}

On this dataset, we show in \Cref{fig:rankplanes} the correlation between the probability of coverage by a random walker and the layer centrality value we compute.
For each company $\alpha$ corresponding to each layer, and for a given $t>t_0$, we compute the random variable $X_{\alpha}(v,t) = \sum_{(t,(u,\alpha),(v,\alpha))\in E_M} (t_{max}-t)$. In other words, the probability to take a plane decrease with time, until it reaches $0$ after a certain amount of time.


Given these probabilities, we compute then the co-variance matrix of those variables, as explained in \Cref{sec:centralities}. The eigenvalues of the covariance matrix corresponds to the centrality score. Notice that, however, just like in graphs, the centrality scores themselves mean little, as it is their relative order that carries importance.

In order to assess the usefulness of our metric, we show in \Cref{fig:rankplanes} the rank of each company compared to the relative coverage of each company by a random (temporal) walker.
In the four subdatasets that we consider, we can see that our centrality is well correlated to the random walker coverage, though however this is especially true for the older subdatasets (1988 and 1995).


We notice that the layer centrality fits less and less with the one of random walker over time. This is due to the fact that the score of each layer in the eigenvector tends to be the same. This means that a lot of companies tends to look like the other ones. We can gess that with the improvement of the study of the market, the carriers have found what are the more interesting routes and concentrates on the same ones. The number of flight has increased a lot (436,951 per month in 1988 to 583,986 in 2019) but probably on the most popular routes rather than to create new connections.

\section{Conclusion}
\label{sec:conclusion}

In this paper, we devise a new formalism that bridges the gap between two recent advances in the state-of-the-art: multilayer graphs and stream graphs.
We propose a new framework that generalizes both objects, and define some elementary notions on it, in order to show its relevance.
Furthermore, we introduce two notions of {\em layer centrality} that capture the relative importance of layers over time.
We experiment on two interaction datasets, of individual contacts and flight information, and show the relevance of the formalism and centralities at capturing subtle patterns in the data.

This work is intended only as a validation for the multilayer stream graph model, and as such it opens numerous perspectives.
The first of them relates to the formalism itself: while the model we define is straightforwardly usable, it can be extended in many ways.
While some of these are straightforward, such as directionality, others require more thorough work, such as ponderation, or proper label utilization.

Another interesting axis depends on the data itself. W many examples of multilayer stream graphs exist in real life, all the relevant information is not typically captured in datasets, typically because the current models cannot use the extra information. We hope this paper serves as a wider call to researchers of many disciplines, to use our model and tailor it to their needs.

%
%

\bibliography{bibliography.bib}
\bibliographystyle{splncs03}
\end{document}